\documentclass[aps,preprint]{revtex4}%
\usepackage{amsfonts}
\usepackage{amsmath}
\usepackage{amssymb}
\usepackage{graphicx}%
\begin{document}
\preprint{ }
\title{Why current-carrying magnetic flux tubes gobble up plasma and become thin as a result}
\author{P. M. Bellan}
\affiliation{MC 128-95 Caltech, Pasadena CA 91125}
\keywords{one two three}
\pacs{96.60.Pb, 52.30.-q, 52.50.Lp, 52.55.Wq, 52.58.Lq, 95.30.Qd }

\begin{abstract}
Suppose an electric current $I$ flows along a magnetic flux tube that has
poloidal flux $\psi$ and radius $a=a(z)$ where $z$ is the axial position along
the flux tube. This current creates a toroidal magnetic field $B_{\phi}.$ It
is shown that, in such a case, non-linear, non-conservative $\mathbf{J\times
B}$ forces accelerate plasma axially from regions of small $a$ to regions of
large $a$ and that this acceleration is proportional to $\partial
I^{2}/\partial z$. Thus, if a current-carrying flux tube is bulged at, say,
$z=0$ and constricted at, say, $z=\pm h$, then plasma will be accelerated from
$z=\pm h$ towards $z=0$ resulting in a situation similar to two water jets
pointed at each other. The ingested plasma convects embedded, frozen-in
toroidal magnetic flux from $z=\pm h$ to $z=0$. The counter-directed flows
collide and stagnate at $z=0$ and in so doing (i) convert their translational
kinetic energy into heat, (ii) increase the plasma density at $z\approx0$, and
(iii) increase the embedded toroidal flux density at $z\approx0$. The increase
in toroidal flux density at $z\approx0$ increases $B_{\phi}$ and hence
increases the magnetic pinch force at $z\approx0$ and so causes a reduction of
\ $a(0)$. Thus, the flux tube develops an axially uniform cross-section, a
decreased volume, an increased density, and an increased temperature. This
model is proposed as a likely hypothesis for the long-standing mystery of why
solar coronal loops are observed to be axially uniform, hot, and bright. It is
furthermore argued that a small number of tail particles bouncing between the
approaching counterstreaming plasma jets should be Fermi accelerated to
extreme energies. Finally, analytic solution of the Grad-Shafranov equation
predicts that a flux tube becomes axially uniform when the ingested plasma
becomes hot and dense enough to have \ $2\mu_{0}n\kappa T/B_{pol}^{2}%
=\ (\mu_{0}Ia(0)/\psi)^{2}/2$; observed coronal loop parameters are in
reasonable agreement with this relationship which is analogous to having
$\beta_{pol}=1$ in a tokamak.

\end{abstract}
\volumeyear{year}
\volumenumber{number}
\issuenumber{number}
\eid{identifier}
\preprint{to appear in Physics of Plasmas}
\accepted{for publication January 14, 2003 }

\startpage{1}
\endpage{ }
\maketitle

\section{\qquad Introduction}

A long standing mystery in solar physics is why solar coronal loops typically
have an axially uniform cross-section \cite{Klimchuk:2000}; i.e., a
filamentary shape. This issue has been made especially pressing by recent
TRACE\ (Transition Region and Coronal Explorer ) spacecraft soft x-ray images
which show a multitude of highly-defined axially uniform loops
\cite{Aschwanden:2000}; for example, see Fig. 1. Axial uniformity of flux
tubes \ is also commonly observed in laboratory experiments, for example, in
recent simulations of solar prominences \cite{Hansen:2001}.%

\begin{figure}
[h]
\begin{center}
\includegraphics[
height=5.5267in,
width=5.5267in
]%
{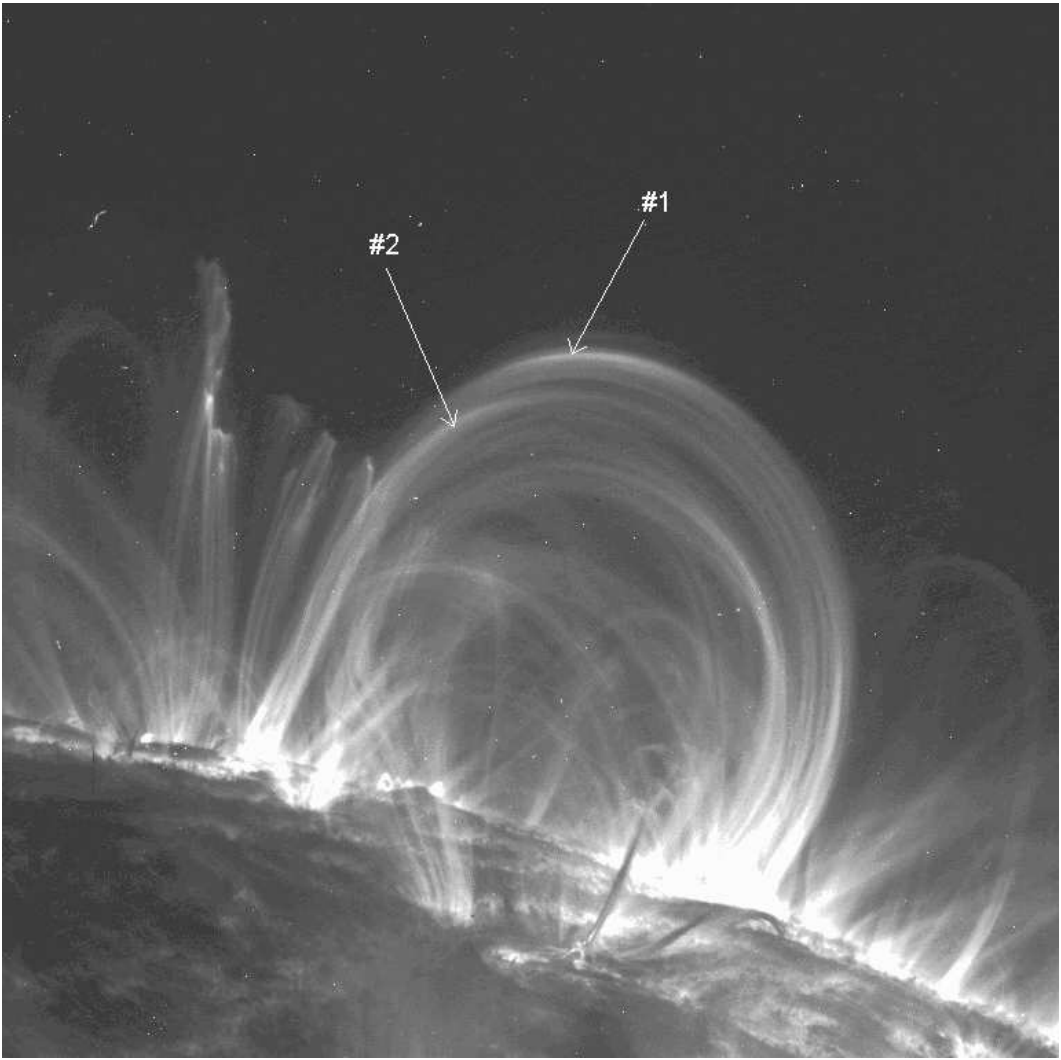}%
\caption{TRACE soft X-ray photo showing axial uniformity of coronal loops
(image courtesy Lockheed Martin Space and Astrophysics Lab). As discussed in
the text, the slight wrapping of flux tube \#1 around flux tube \#2 indicates
that net currents flow along the flux tubes.}%
\end{center}
\end{figure}

This paper argues that axial uniformity is the result of a rather complex
sequence of events which occur whenever an electric current $I$ is made to
flow along an initially axially non-uniform, current-free, axisymmetric
magnetic flux tube (a process corresponding to injection of magnetic helicity
into the flux tube). The sequence of events occurs even when $I$ is modest,
i.e., even when the flux tube is only slightly twisted.%

\begin{figure}
[h]
\begin{center}
\includegraphics[
height=5.3009in,
width=1.9195in
]%
{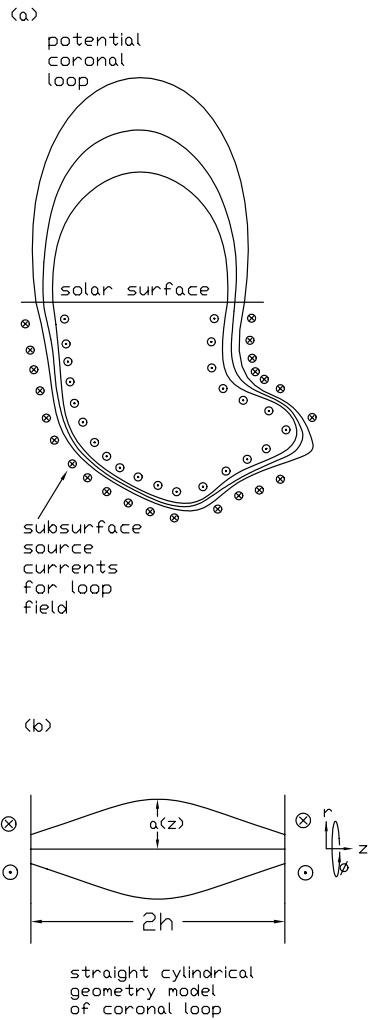}%
\caption{(a) Potential (i.e., current-free) coronal loop above solar surface
with corresponding subsurface field and source currents. The cross-section of
the loop is largest at the top of the loop where the magnetic field is
weakest; (b) straight cylindrical representation of coronal loop used in
model.}%
\end{center}
\end{figure}

The typical arched shape of coronal loops is shown schematically in Fig. 2a;
to make the analysis tractable we will assume that the loop is straight as
sketched in Fig. 2b. However, in order to retain an important aspect of the
arched shape, we will allow the length of the straight loop to vary to take
into account possible variability in the length of the arched loop. The Fig.
2b geometry will be characterized by a straight cylindrical coordinate system
$\{r,\phi,z\}$ where $z$ refers to the direction along the loop axis, $\phi$
is the azimuthal direction about the axis, and $r$ is the distance from the
axis. The $\phi$ direction is called the toroidal direction and the $r,z$
directions are called poloidal. Flux coordinates will also be used when
appropriate. This poloidal/toroidal nomenclature is formally the same as that
used for tokamaks, but the configuration should not be confused with a tokamak
as there are no closed poloidal field lines. The current $I$ will be assumed
to be relatively small so that the poloidal field magnitude $\sim B_{z}$ is
always much larger than $B_{\phi}$ in which case the flux tube is only
slightly twisted. We note that this straight cylindrical approximation of flux
tube geometry has been used in many previous studies of flux tube equilibria,
especially force-free equilibria (for example, see
Refs.\cite{Parker:1979,Browning:1983,Zweibel:1985,Browning:1989,Lothian:1989,VanderLinden:1999}%
). However, the analysis presented here differs substantively from these
previous studies because our analysis does not begin by assuming existence of
an equilibrium. Instead, our analysis characterizes the dynamics that lead to
an equilibrium and shows how the resulting equilibrium is intimately related
to these dynamics. Furthermore, our analysis takes into account the
non-force-free aspects of the equilibrium (i.e., finite pressure gradients)
and shows that these finite $\beta$ aspects are of vital importance to the
axial uniformity of the equilibrium.

Because of the assumed axisymmetry in Fig. 2b, the magnetic field can be
expressed as
\begin{equation}
\mathbf{B=}\frac{1}{2\pi}\left(  \nabla\psi\times\nabla\phi+\mu_{0}I\nabla
\phi\right)  \label{defB}%
\end{equation}
where $\nabla\phi=\hat{\phi}/r$ and
\begin{equation}
\psi(r,z,t)=\int_{0}^{r}B_{z}(r^{\prime},z,t)2\pi r^{\prime}dr^{\prime}
\label{flux}%
\end{equation}
is the poloidal flux. Axial non-uniformity corresponds to having $\partial
\psi/\partial z\neq0\ $\ and axial bulging corresponds to having $\psi
^{-1}\partial^{2}\psi/\partial z^{2}>0.$ The current $I$ is similarly given
as
\begin{equation}
I(r,z,t)=\int_{0}^{r}J_{z}(r^{\prime},z,t)2\pi r^{\prime}dr^{\prime} \label{I}%
\end{equation}
and is related to the toroidal field by Ampere's law, i.e., $\mu_{0}I=2\pi
rB_{\phi}.$ The current rise time $\tau$ is assumed to be sufficiently slow
that Alfv\'{e}n wave propagation effects are unimportant i.e., it is assumed
that $v_{A}\tau>>h$ where $2h$ is the length of the flux tube. The current
thus flows as in an electric circuit so that there are no retarded time or
radiation effects. Taking the curl of Eq.(\ref{defB}) shows that the current
density associated with the magnetic field is
\begin{equation}
\mathbf{J=-}\frac{r^{2}\nabla\phi}{2\pi\mu_{0}}\nabla\cdot\left(  \frac
{1}{r^{2}}\nabla\psi\right)  +\frac{1}{2\pi}\nabla I\times\nabla\phi.
\label{J}%
\end{equation}

Controversy exists regarding the properties of $I(r,z)$ external to the
current-carrying flux tube$.\ $ Some authors \cite{Parker:1996,Linton:2001}
argue that $I$ must vanish outside the flux tube while others
\cite{Mok:1997,Longcope:2000,Chen:2001,Melrose:1995} argue that $I$ should be
finite outside. If one insists that $I$ vanishes outside the flux tube so that
there is no net current in a flux tube, the flux tube acts like a coaxial
cable (i.e., a center conductor sheathed by a coaxial outer conductor carrying
equal and opposite current). This neutralized current configuration has
$B_{\phi}=0$ external to the flux tube and so cannot produce magnetic forces
on external currents. Adjacent neutralized flux tubes thus cannot exert forces
on each other and so will not mutually interact, just like adjacent coaxial
cables will not mutually interact. Furthermore, sections at different axial
positions along the length of a neutralized flux tube cannot interact via
magnetic forces. The lack of interaction between sections at different axial
positions along the length of a neutralized flux tube means that such a flux
tube cannot undergo a kink instability since a kink instability involves
relaxing to the state of lowest mutual interaction energy between loop
segments (similarly, a coaxial cable will not undergo a kink instability since
there is no force between axially separated segments).

In contrast, having a net current (i.e., being non-neutralized) implies
existence of an external potential magnetic field $B_{\phi}$ $\sim I/r$
outside of the flux tube just like the magnetic field external to an ordinary
current-carrying wire. An individual flux tube with net current can kink
since, if its axis bends, there will be forces between sections at different
axial positions. Two flux tubes, each carrying net current,\ will experience
mutual interaction forces due to the $B_{\phi}$ of one flux tube interacting
with the current of the other flux tube. Thus, two adjacent flux tubes each
with net current will tend to wrap around each other as shown in Fig. 2a of
Ref.\cite{Lau:1996} since the axis of each flux tube will be affected by the
$B_{\phi}$ of its neighbor. Since we are assuming here that $B_{\phi}<<B_{z}$
this wrapping will be very slight. Examination of the loop structures \#1 and
\#2 denoted by arrows in Fig.1 show that these two structures do indeed wrap
around each other slightly (on the left, loop \#1 is to the rear of loop \#2;
on the right, the two loops appear to be in the plane of the photo).

Based on the observations that kinks do occur in solar structures and that
coronal loops do show evidence of wrapping we will assume in this paper that
net current does flow in a flux tube, i.e., that the current is
non-neutralized. This assumption is additionally supported by recent work by
Feldman \cite{Feldman:2002} and by Wheatland\cite{Wheatland:2000} who, after
careful analysis of a variety of observational evidence, have concluded that
net currents do flow.

The assumption of net current means that we are allowing flux tubes to
interact with each other via magnetic forces. However, since the effect of
this interaction is to alter the \ three dimensional locus of a flux tube
axis, and since we are invoking the straight axis approximation (i.e. using
Fig. 2b to represent Fig. 2a), we are removing from consideration the
evolution of the three dimensional locus of the flux tube axis. The straight
axis approximation is thus analogous to a kinematics problem where one works
in the center of mass frame of a body and so removes from consideration
external body forces that change the location of the center of mass.

The dynamics of the configuration are governed by the combination of the
magnetohydrodynamic (MHD)\ equation of motion%
\begin{equation}
\rho\frac{d\mathbf{U}}{dt}=\mathbf{J\times B}-\nabla P \label{motion}%
\end{equation}
and the induction equation%
\begin{equation}
\frac{\partial\mathbf{B}}{\partial t}=\nabla\times\left(  \mathbf{U\times
B}\right)  \label{induction}%
\end{equation}
where the latter is obtained from the curl of the ideal MHD\ Ohm's law%
\begin{equation}
\mathbf{E+U\times B=}0. \label{Ohm}%
\end{equation}

The flux tube ends are at $z=\pm h$ and the flux tube middle is at $z=0.$ The
flux tube is assumed to be initially current-free so that initially $I=0.$ The
field in the flux tube is thus initially potential (i.e., $\nabla
\times\mathbf{B=}0$ initially) and the source currents generating this
potential field are external to the flux loop and, as sketched in Fig. 2a, are
assumed to be below the solar surface. Thus the initially potential coronal
loop sketched in Fig. 2a will have a magnetic field that is stronger at the
footpoints than at the arch top because the arch top is further from the
source currents. This means that the initial current-free flux tube will be
bulged at the top because the magnetic field is weaker there. In the straight
geometry representation given by Fig.2b, the bulging at the arch top
corresponds to having the initially potential poloidal field stronger at
$z=\pm h$ than at $z=0$ and the flux tube diameter larger at $z=0$ than at
$z=\pm h.$

The sequence of events that occurs when the current is ramped up to a
steady-state will be shown to consist of the following three stages:

\begin{enumerate}
\item The first stage consists of a twisting of the magnetic field about the
$z$ axis in Fig. 2b together with an associated transient toroidal plasma
velocity $U_{\phi}$. This stage is incompressible and maintains the flux tube
poloidal profile, i.e., $\psi(r,z)$ is unchanged and the flux tube remains
bulged. The velocity $U_{\phi}$ is proportional to $z\partial I/\partial t$
and the toroidal acceleration is proportional to $z\partial^{2}I/\partial
t^{2}.$

\item The second stage involves generation of axial plasma flows. These flows
go from $z=\pm h$ where the flux tube diameter is small to $z=0$ where the
diameter is large. The flows are driven by a $z$-directed force which is
proportional to $-\partial I^{2}/\partial z.$ This axial force is a nonlinear
function of $I$ in contrast to the first stage toroidal acceleration which is
a linear function of $I.$

\item The third stage involves stagnation of the converging flows at $z=0$
resulting in plasma heating as the flow kinetic energy is converted into heat
(thermalized). There is also an accumulation of convected \textit{toroidal}
flux at $z=0$ which leads to an enhancement of the pinch force at $z=0$. The
enhanced pinch force squeezes the flux tube diameter at $z=0$ so that the flux
tube approaches axial uniformity, i.e., $\partial\psi/\partial z\rightarrow0.$
Ultimately, an axially uniform flux \ tube loaded with hot plasma results.
\end{enumerate}

\qquad

\section{ Lack of equilibrium for arbitrarily specified magnetic fields}

Arbitrarily specified magnetic fields do not, in general, have associated
MHD\ equilibria, i.e., in general, no pressure $P(\mathbf{r)}$ exists which
satisfies
\begin{equation}
\mathbf{J\times B}=\nabla P \label{equiln}%
\end{equation}
for arbitrarily specified $\mathbf{B(r).}$ The essential physics underlying
this assertion is that $\nabla\times\nabla P$ is identically zero whereas
$\nabla\times\left(  \mathbf{J\times B}\right)  $ is not necessarily zero,
i.e., $\nabla P$ is always a conservative force whereas $\mathbf{J\times B}$
is in general non-conservative. A non-conservative force has an associated
torque and since a pressure gradient cannot balance a torque, no equilibrium
is possible when $\nabla\times\left(  \mathbf{J\times B}\right)  $ is finite.%

\begin{figure}
[h]
\begin{center}
\includegraphics[
height=1.1108in,
width=3.6164in
]%
{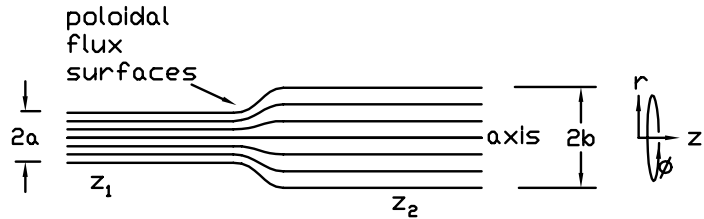}%
\caption{Poloidal flux surfaces of a magnetic field configuration for which no
finite pressure gradient MHD\ equilibrium is possible. }%
\end{center}
\end{figure}

As a specific example that equilibria do not exist for arbitrarily specified
magnetic field configurations, consider the simple situation sketched in Fig.
3. We assert that no MHD\ equilibrium is possible for this configuration,
i.e., this configuration cannot satisfy Eq. (\ref{equiln}). This assertion is
established by calculating radial pressure balance at axial locations
$\ \ z_{1}$ and $z_{2}$ where the respective flux tube radii are $a$ and
$b\ $with $b>a.$ If satisfied, Eq. (\ref{equiln}) imposes the requirement
$\mathbf{B\cdot}\nabla P=0$ which means that $\ $the pressure must be the same
everywhere on a field line. \qquad

The poloidal component of the magnetic field is assumed to be straight,
uniform, and in the $z$ direction at both $z_{1}$ and $z_{2}.$ Because the
field lines are more densely packed at $z_{1}$ than at $z_{2},$ the magnetic
field is such that $B_{z}(z_{1})>B_{z}(z_{2}).$ In addition, since the
poloidal magnetic field is uniform at \ $z_{1}$ and $z_{2},$ the toroidal
current vanishes at both $z_{1}$ and $z_{2}$ [in fact, uniformity of poloidal
field is more than is needed for the toroidal current to vanish as the most
general requirement for the toroidal current to vanish is to have $\nabla
\cdot\left(  r^{-2}\nabla\psi\right)  =0$].

Using $J_{\phi}=0$ the radial component of Eq.(\ref{equiln}) is therefore
\begin{equation}
\partial P/\partial r=-J_{z}B_{\phi}=-\frac{\mu_{0}I}{\left(  2\pi r\right)
^{2}}\frac{\partial I}{\partial r}. \label{dpdr}%
\end{equation}
However, the toroidal component of Eq.(\ref{equiln}) gives
\begin{equation}
\left(  \nabla I\times\nabla\phi\right)  \times\left(  \nabla\psi\times
\nabla\phi\right)  =0 \label{torequiln}%
\end{equation}
which implies that $I=I(\psi).$ Since the poloidal magnetic field is straight
and uniform at both $\ z_{1}\ $and $z_{2}$, the poloidal flux function must
have the form $\psi=\psi_{0}r^{2}/a^{2}$ $\ $at $\ z_{1}$ and $\psi=\psi
_{0}r^{2}/b^{2}$ at $\ z_{2}$ where $\psi_{0}$ is the flux on the surface for
which $P$ vanishes and $a,b$ are the respective radii of these flux surfaces
at $z_{1}$ and $z_{2}.$ The simplest non-trivial possibility for $I=I(\psi)$
is to assume that $I$ is a linear function of $\psi$ so
\begin{equation}
\mu_{0}I=\alpha\psi. \label{muIphi}%
\end{equation}
Radial pressure balance at $z=z_{1}$ thus becomes%
\begin{equation}
\partial P/\partial r=-\frac{\alpha^{2}\psi}{\left(  2\pi r\right)  ^{2}%
\mu_{0}}\frac{\partial\psi}{\partial r}=-\frac{\alpha^{2}\psi_{0}^{2}}%
{2\pi^{2}\mu_{0}}\frac{r}{a^{4}} \label{radialpressurebal}%
\end{equation}
which can be integrated to give the on-axis pressure at $z_{1}$ to be%
\begin{equation}
P(0,z_{1})=\int_{a}^{0}dr\partial P/\partial r=\int_{0}^{a}dr\frac{\alpha
^{2}\psi_{0}^{2}}{2\pi^{2}\mu_{0}}\frac{r}{a^{4}}=\frac{\alpha^{2}\psi_{0}%
^{2}}{4\pi^{2}a^{2}\ \mu_{0}}. \label{onaxisP}%
\end{equation}
However, a similar evaluation of the on-axis pressure at $z=z_{2}$ gives%
\begin{equation}
P(0,z_{2})=\frac{\alpha^{2}\psi_{0}^{2}}{4\pi^{2}b^{2}\mu_{0}}
\label{on-axisP2}%
\end{equation}
which is smaller than the on-axis pressure at $z=z_{1}$ since $b>a$. Thus, the
pressure is \textit{not} uniform along the on-axis field line and so the
requirement $\mathbf{B}\cdot\nabla P=0$ is violated on the $z$-axis. Equation
(\ref{equiln}) is therefore not satisfied and so equilibrium does not exist.
Because there is radial pressure balance but not axial pressure balance, we
might expect flows to be driven from $z_{1}$ to $z_{2}.$ The situation is
analogous to squeezing the end of a toothpaste tube\ \ and having toothpaste
squirt out the mouth of the tube.

\section{General case using flux coordinates}

Let us now return to the problem of a coronal loop which is initially
current-free and bulging at $z=0$, but then has an externally driven current
$I$ slowly ramped up to a constant value. We define a right-handed orthogonal
coordinate system based on the poloidal flux coordinates. The unit vectors of
this system are%
\begin{align}
\mathbf{e}_{\psi}  &  =\frac{\nabla\psi}{|\nabla\psi|}\nonumber\\
\mathbf{e}_{\phi}  &  =\hat{\phi}\nonumber\\
\mathbf{e}_{Bpol}  &  =\frac{\nabla\psi\times\nabla\phi}{|\nabla\psi
\times\nabla\phi|} \label{unitvectors}%
\end{align}
so that $\mathbf{e}_{\psi}\rightarrow$ $\hat{r}$ and $\mathbf{e}_{Bpol}$
$\rightarrow\hat{z}$ if the field lines are straight and axial. \ The form of
$\mathbf{e}_{\psi}$ shows that $B_{\psi}=\mathbf{B\cdot e}_{\psi}$ is zero by
definition; i.e., the magnetic field can never have a component in the
direction of $\nabla\psi.$

The current density can be decomposed into the components
\begin{equation}
J_{\psi}=\frac{1}{2\pi}\nabla I\times\nabla\phi\cdot\frac{\nabla\psi}%
{|\nabla\psi|} \label{Jpsi}%
\end{equation}%
\begin{equation}
J_{\phi}=\mathbf{-}\frac{r}{2\pi\mu_{0}}\nabla\cdot\left(  \frac{1}{r^{2}%
}\nabla\psi\right)  \label{Jphi}%
\end{equation}%
\begin{equation}
J_{Bpol}=\frac{1}{2\pi r}\frac{\nabla I\cdot\nabla\psi}{|\nabla\psi|}.
\label{JBpol}%
\end{equation}
We now argue that $J_{\psi},J_{\phi},$ and $J_{Bpol}$ each have distinctive physics.

The component $J_{\psi}$ flows normal to flux surfaces and provides the torque
that causes the plasma to rotate toroidally. This current can only be
transient and is identified as the polarization current \cite{Chen:1984}.
Polarization current can be thought of as being an essentially dependent
quantity; that is, one first determines the amount of toroidal acceleration
using an analysis that does not involve the equation of motion, and then one
inserts this acceleration into the equation of motion to calculate the
required polarization current. The reason for this inferior status of the
toroidal component of the equation of motion is that the toroidal symmetry of
the system provides a strong constraint on the dynamics. From an MHD\ point of
view, the toroidal symmetry means that no toroidal pressure gradient can exist
and also no toroidal electrostatic electric field can exist. From a particle
Hamiltonian point of view, this symmetry means that the maximum excursion
particles can make from a flux surface is no more than a poloidal Larmor
radius \cite{Bellan:2000}, a microscopic length. The localization of particles
to the vicinity of a flux surface means that there cannot be any sustained net
current density in the direction normal to a flux surface and so $J_{\psi}$ is
highly constrained. All that is allowed is a short-lived transient $J_{\psi
}^{\ }$ having zero time-average; i.e., $J_{\psi}$ can only be an AC current.
The slight bobbing back and forth of particles off of a flux surface
constitutes the polarization current. Thus the plasma acts like a capacitor in
the direction normal to the flux surfaces, but like a wire in the direction
along the flux surfaces. As is well known\cite{Chen:1984a}, the dielectric
constant of the plasma \textquotedblleft capacitor" is given by $\mu_{0}\rho
c^{2}/B^{2}=c^{2}/v_{A}^{2}>>1.$ Polarization currents have an associated
polarization electric field normal to the flux surface resulting from the
particles making their small excursions from their nominal flux surface.

The current $I$ is assumed to be generated by some sub-surface dynamo and so
its time-dependence is a prescribed quantity. This time dependence is assumed
to be such that $I$ increases smoothly from zero to some finite value in a
characteristic time $\tau>>h/v_{A}.$ This smooth increase can be represented
by the characteristic time profile%
\begin{equation}
I(t)=\frac{\left(  \tanh(t/\tau)+1\right)  }{2}I_{0} \label{tanh}%
\end{equation}
so that $I=0$ for $t<<-\tau$ and $I=I_{0}$ for $t>>\tau.$ It should be noted
that $\partial I/\partial t\sim1/\cosh^{2}(t/\tau)$ which has its maximum at
$t=0$ and that $\partial^{2}I/\partial t^{2}\sim-\tanh\left(  t/\tau\right)
/\cosh^{2}(t/\tau)$ which is positive for $t$ slightly before $t=0$ and
negative for $t$ slightly after $t=0$ and then otherwise zero.%

\begin{figure}
[h]
\begin{center}
\includegraphics[
height=3.1067in,
width=3.6106in
]%
{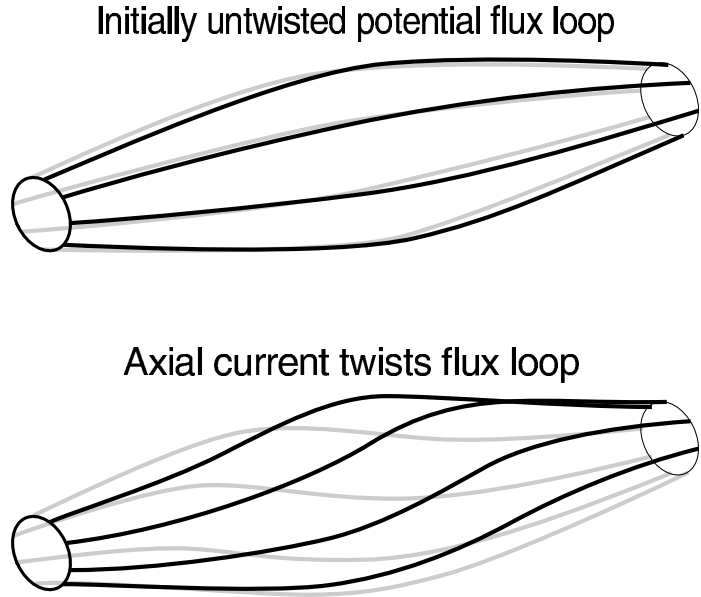}%
\caption{First stage dynamics: Top shows flux tube when $I=0.$ Bottom shows
flux tube when $I$ is finite, but $\psi(r,z)$ is not yet changed from its
initial potential value. The evolution from the top to the bottom situation
involves toroidal rotation by a finite amount, but no poloidal flows.}%
\end{center}
\end{figure}

\subsection{First stage (ramp-up)}

At the beginning of the first stage $I$ is zero and the flux tube is untwisted
as shown in\ the top sketch of Fig. 4. \ The toroidal component of the
induction equation is
\begin{equation}
\frac{\partial B_{\phi}}{\partial t}=\ r\mathbf{B}_{pol}\mathbf{\cdot}%
\nabla\left(  \frac{U_{\phi}}{r}\right)  -r\mathbf{U}_{pol}\mathbf{\cdot
}\nabla\left(  \frac{B_{\phi}}{r}\right)  -B_{\phi}\nabla\cdot\mathbf{U}%
_{pol}\ \label{torBphi}%
\end{equation}
and the toroidal component of Ohm's law can be expressed as%
\begin{equation}
\frac{\partial\psi}{\partial t}\mathbf{+U}_{pol}\mathbf{\cdot}\nabla
\psi\mathbf{=}0. \label{psiconv}%
\end{equation}
We will show that $\mathbf{U}_{pol}$ results from a weak non-linear force and
so is negligible at times $t\sim0$ because there has not been enough time for
a significant $\mathbf{U}_{pol}$ to develop. In contrast, $U_{\phi}$ is
proportional to $\partial I/\partial t$ and so $U_{\phi}$ is large at
$t\sim0\ $when $\partial I/\partial t$ is at its maximum. However $U_{\phi}$
goes to zero at large times when $\partial I/\partial t\rightarrow0.$ Thus,
the time when $U_{\phi}$ is finite precedes the time when $\mathbf{U}_{pol}$
is finite. The first stage characterizes the time $t\sim0$ when $\mathbf{U}%
_{pol}$ is negligible, $\psi$ is unchanged from its initial potential state,
$J_{\phi}$ is negligible, $\ $and $U_{\phi}$ is transiently finite.

Letting $s$ denote the distance along the poloidal field from the $z=0$ plane
and taking into account that $\mathbf{U}_{pol}\simeq0$, Eq.(\ref{torBphi}) may
therefore be approximated at $t\sim0$ as%
\begin{equation}
\frac{\partial B_{\phi}}{\partial t}\simeq B_{pol}\mathbf{\ }\frac
{\mathbf{\partial}U_{\phi}}{\partial s}\ \label{dBphidt}%
\end{equation}
which may be integrated with respect to $s$ to give%
\begin{equation}
U_{\phi}\simeq\frac{s}{B_{pol}}\frac{\partial B_{\phi}}{\partial t}=\frac
{\mu_{0}s}{2\pi B_{pol}r}\frac{\partial I\ }{\partial t}. \label{solveUphi}%
\end{equation}
The finite toroidal displacement $r\Delta\phi=\int U_{\phi}dt$ is proportional
to $s$ and so gives a twisting up of the flux tube as shown in the bottom
sketch of Fig. 4. In fact this twisting motion is such that
\begin{equation}
\frac{r\Delta\phi}{B_{\phi}}=\ \frac{\ s}{\ B_{pol}\ }\ \label{twist}%
\end{equation}
showing that the plasma twist is just what is required to keep the plasma
frozen to the twisting magnetic field.

Equation (\ref{solveUphi}) and Eq.(\ref{Ohm}) together imply the existence of
an electric field in the $\nabla\psi$ direction
\begin{equation}
E_{\psi}=-U_{\phi}B_{pol}=-\frac{\mu_{0}s}{2\pi r}\frac{\partial I\ }{\partial
t};\ \label{Epsi}%
\end{equation}
this is the polarization electric field. The toroidal component of the
equation of motion is
\begin{equation}
\rho\frac{\partial U_{\phi}}{\partial t}\simeq-J_{\psi}B_{pol}
\label{tormotion}%
\end{equation}
since $B_{\psi}=0$. Thus, the current normal to a flux surface is
\begin{equation}
J_{\psi}=-\frac{\rho}{B_{pol}}\frac{\partial U_{\phi}}{\partial t}=\frac{\rho
}{B_{pol}^{2}}\frac{\partial E_{\psi}}{\partial t}=-\frac{\rho}{B_{pol}^{2}%
}\frac{\mu_{0}s}{2\pi r}\frac{\partial^{2}I\ }{\partial t^{2}}.
\label{solveJpsi}%
\end{equation}
Equation (\ref{solveJpsi}) clearly shows that $J_{\psi}$ is\ indeed the
polarization current and that $J_{\psi}$ is essentially a dependent quantity
since it is proportional to $\partial^{2}I/\partial t^{2}.$ The polarization
current is transient and, for positive $s,$ is first negative and then
positive (and vice versa for negative $s$). Both $U_{\phi}$ and the
polarization current $J_{\psi}$ vanish when $I$ is in steady state. The chain
of dependence is such that the induction equation first determines $U_{\phi}$
which then determines $J_{\psi}\ $via the equation of motion. For a long, thin
flux tube, $s\simeq z,$ $B_{pol}\simeq B_{z}$, and the $\psi$ direction is
approximately the $r$ direction. For $t>>\tau$, the poloidal current $I$ is in
steady state and so $U_{\phi}=0\ $for $t>>\tau.$

\bigskip

\subsection{Second stage (steady $I,$ development of finite $\mathbf{U}_{pol}%
$)}

Since $I$ is constant in the second stage, $U_{\phi}\ $and $J_{\psi}$ are both
zero. Equation (\ref{Jpsi}) shows that $J_{\psi}=0$ $\ $implies $I=I(\psi
)\ $in which case surfaces of constant $\psi(r,z)$ are also surfaces of
constant $I(r,z)$. At the beginning of the second stage, $\mathbf{U}_{pol}$
has not yet developed and so $\psi(r,z)$ is assumed to be unchanged from its
initial potential (vacuum) state, i.e., the poloidal profile of the flux
surfaces is not yet deformed. Thus, $J_{\phi}=\mathbf{-}r\left(  2\pi\mu
_{0}\right)  ^{-1}\nabla\cdot\left(  r^{-2}\nabla\psi\right)  =0$ at the
beginning of the second stage.

We now consider the dynamics. The magnetic\ force can be decomposed into
toroidal and poloidal components as follows:
\begin{equation}
\mathbf{J\times B}=\mathbf{J}_{pol}\mathbf{\times B}_{pol}+\mathbf{J}%
_{pol}\mathbf{\times B}_{tor}+\mathbf{J}_{tor}\mathbf{\times B}_{pol}.
\label{JxB}%
\end{equation}
However, $\mathbf{J}_{pol}\mathbf{\times B}_{pol}$ $=\mathbf{J}_{\psi
}\mathbf{\times B}_{pol}=0$ and $\mathbf{J}_{tor}=J_{\phi}\hat{\phi}=0$ so
that the magnetic\ force at the beginning of the second stage reduces to
\begin{equation}
\mathbf{J\times B}=\mathbf{J}_{pol}\mathbf{\times B}_{tor}=\frac{1}{2\pi
}\left(  \nabla I\times\nabla\phi\right)  \times\frac{\mu_{0}I}{2\pi}%
\nabla\phi=-\frac{\mu_{0}}{8\pi^{2}r^{2}}\nabla I^{2}. \label{netJxB}%
\end{equation}
Since the curl of the magnetic force given in Eq.(\ref{netJxB}) is non-zero,
it is impossible for a pressure gradient to balance the magnetic force at this
stage. The $z$ component of the magnetic force is%
\begin{equation}
\ \left(  \mathbf{J\times B}\right)  _{z}=-\frac{\mu_{0}}{8\pi^{2}r^{2}}%
\frac{\partial I^{2}}{\partial z} \label{Fz}%
\end{equation}
which is independent of the sign of $I$, nonlinear in $I,$ and such as to
accelerate plasma from regions where the diameter of the current channel is
small to regions where the diameter is large. In the case of a flux tube which
is bulged in the middle, the force given in Eq.(\ref{Fz}) will accelerate
plasma axially from $z=\pm h$ towards $z=0$.

The force given in Eq.(\ref{Fz}) vanishes at $r=0$ since $I\sim r^{2}$ for
small $r.$ However, axially localized radial force balances will quickly
develop between the magnetic force and the radial pressure gradient as
discussed in Section II. The resulting radial force balance will produce an
axially non-uniform pressure and so there will also be an axial force due to
the axial pressure gradient.

Specifically, radial pressure balance means that
\begin{equation}
\frac{\partial P}{\partial r}=-J_{z}B_{\phi}=-\frac{\mu_{0}}{8\pi^{2}r^{2}%
}\frac{\partial I^{2}}{\partial r}. \label{dPdr}%
\end{equation}
This radial pressure balance equation is not integrable for arbitrary
$I(r,z)$, but to get an idea for the general behavior we make the assumption
that $I\sim r^{2}$ which is integrable. Thus, assuming $I(r,z)=\left(
r/a(z)\right)  ^{2}I_{0}$ where $I_{0}$ is the total current flowing in the
flux tube of radius $a(z),$ Eq.(\ref{dPdr}) can be integrated to give%
\begin{equation}
P(r,z)=\frac{\mu_{0}I_{0}^{2}}{4\pi^{2}a^{2}\ \ }\left(  1-\frac{r^{2}}{a^{2}%
}\right)  \label{prz}%
\end{equation}
so that
\begin{equation}
-\frac{\partial P}{\partial z}=\frac{\mu_{0}I_{0}^{2}}{\pi^{2}a^{3}%
\ \ }\left(  \frac{1}{2}-\frac{r^{2}}{a^{2}}\right)  \frac{\partial
a}{\partial z} \label{dPdz}%
\end{equation}
is the axial force due to the axial non-uniformity of the pressure.

Using $I(r,z)=I_{0}r^{2}/a^{2}$ to estimate the axial component of the
magnetic force gives%
\begin{equation}
\left(  \mathbf{J\times B}\right)  _{z}=-\frac{\mu_{0}}{8\pi^{2}r^{2}}%
\frac{\partial I^{2}}{\partial z}=\ \frac{\mu_{0}I_{0}^{2}}{2\pi^{2}}%
\ \frac{r^{2}}{a^{5}}\frac{\partial a}{\partial z}. \label{mag-z}%
\end{equation}

The total force in the $z$ direction for this case is thus
\begin{align}
F_{z}  &  =\ \left(  \mathbf{J\times B}\right)  _{z}-\frac{\partial
P}{\partial z}\nonumber\\
&  =\ \frac{\mu_{0}I_{0}^{2}}{2\pi^{2}a^{3}\ \ }\left(  1-\frac{r^{2}}%
{\ a^{2}}\right)  \frac{\partial a}{\partial z}. \label{totalFz}%
\end{align}
This total force is peaked on the axis and has magnitude
\begin{equation}
F_{z}\sim-\frac{\partial}{\partial z}\left(  \left[  \frac{B_{\phi}^{2}}%
{\mu_{0}}\right]  _{r=a}\right)  .\ \label{Fztotal}%
\end{equation}
This force will result in axial flows from $z=\pm h$ to $z=0$ with velocities
that are of the order of $B_{\phi}(r=a,z=\pm h)/\sqrt{\mu_{0}\rho}.$ Because
$J_{\phi}=0$, the behavior is essentially identical to the situation where
$\psi=0$ and so the flow acceleration mechanism is similar to that discussed
in Refs. \cite{Maeker:1955}-\cite{Bellan:1992} which consider MHD\ arc-jets
for the situation of purely toroidal magnetic fields.

\bigskip%
\begin{figure}
[h]
\begin{center}
\includegraphics[
height=2.8277in,
width=6.2382in
]%
{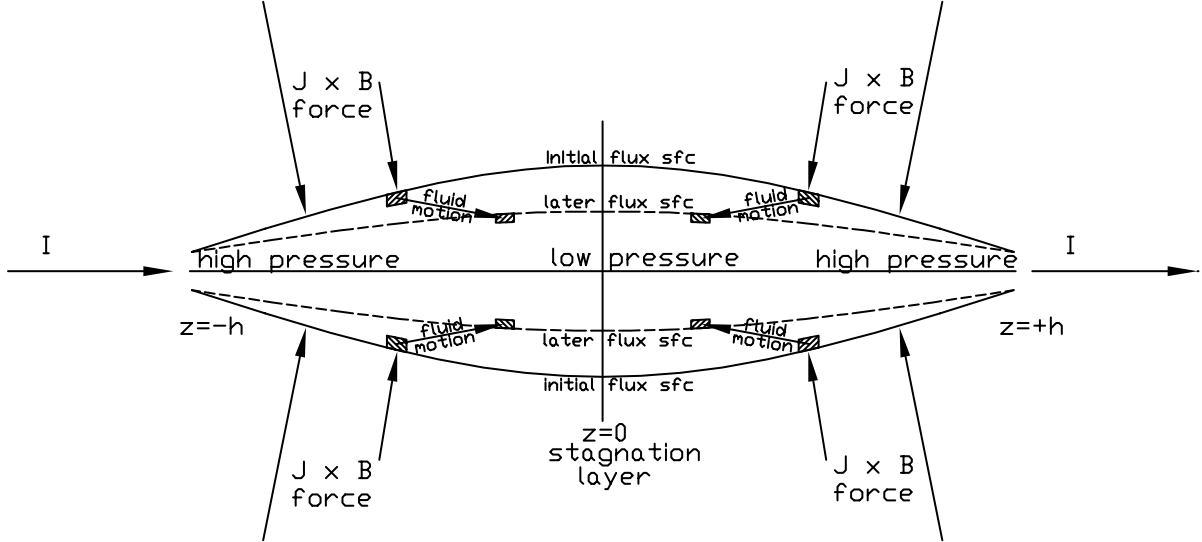}%
\caption{Evolution of flux tube due to effect of $\mathbf{J\times B}$ force
when $I$ is constant. Solid line shows initial constant $\psi(r,z)$ flux
surface, dashed line shows same flux surface at a later time. Fluid elements
(shown hatched) are pushed axially towards $z=0$ while staying on the same
constant $\psi$ surface. The resulting fluid flows collide at the $z=0$ plane
where they thermalize their directed kinetic energy. Toroidal flux and mass
accumulate at the $z=0$ layer also. The accumulation of toroidal flux at $z=0$
increases $B_{\phi}$ there and so pinches down the flux tube diameter causing
the flux tube to become axially uniform.}%
\end{center}
\end{figure}

\subsection{Third stage (convection of toroidal flux, fluid stagnation,
heating, compression)}

The force given by Eq.(\ref{netJxB}) has a finite curl and so cannot be
balanced by a pressure gradient so long as the original potential profile of
$\psi(r,z)$ is maintained. Thus, the only way for an equilibrium to develop is
for the profile of $\psi(r,z)$ to change. This is clearly evident from the
discussion in Section III(B): Eq.(\ref{totalFz}) shows that axial equilibrium
can only occur if $\partial a/\partial z\rightarrow0$.

Attainment of equilibrium involves several inter-related hydrodynamic,
magnetic, and thermodynamic phenomena which are shown schematically in Fig. 5.
The solid lines in Fig. 5 show a constant $\psi$ surface at an early time and
the dashed lines show this constant $\psi$ surface at a later time. Typical
fluid elements are shown as hatched parallelograms (cross-sections of toroids)
and the motion of these elements is seen to consist of both axial and radial
motion such that each fluid element stays on its own constant $\psi$ surface.
The non-conservative nature of the $\mathbf{J\times B}$ force is shown in Fig.
5 by the longer length at larger $|z|$ of the arrows representing
$\mathbf{J\times B}$. The axial motion corresponds to plasma flows which
ingest plasma at $z=\pm h$, travel towards~$z=0,$ and then converge and
stagnate at $z=0\ $like two water jets pointed at each other. The stagnation
converts the flow kinetic energy into heat and simultaneously increases the
plasma density at $z=0$ as plasma accumulates there. Thus pressure increases
at $z=0$. Furthermore, toroidal flux embedded in the plasma is convected by
the axial flows and so there will also be an accumulation of toroidal flux at
$z=0$. This means that the density of the toroidal flux will also increase at
$z=0,$ i.e., $B_{\phi}$ will increase in the vicinity of $z=0.$

The increase in $B_{\phi}$ can be established more rigorously by considering
Eq.(\ref{torBphi}) in the vicinity of $z=0$ and taking into account that
(i)\ $U_{\phi}=0$ since $I$ is constant, (ii) $\mathbf{U}_{pol}\rightarrow0$
at $z=0$ since the flow stagnates at $z=0,$ and (iii) $\nabla\cdot
\mathbf{U}_{pol}<0$ near $z=0$ since the flows are converging at $z=0.$ Thus,
\ Eq.(\ref{torBphi}) in the vicinity of $z=0\,$\ reduces to%

\begin{equation}
\frac{\partial B_{\phi}}{\partial t}\simeq\ -B_{\phi}\nabla\cdot
\mathbf{U}_{pol} \label{dBphidt2}%
\end{equation}
which shows that $B_{\phi}$ must increase since $\nabla\cdot\mathbf{U}%
_{pol}<0$ (we note that amplification of a magnetic field by a converging flow
has previously been discussed in Ref.\cite{Polygiannakis:1999} but has not
otherwise received much attention). In the vicinity of the stagnation layer at
$z=0$, the continuity equation reduces to%
\begin{equation}
\frac{1}{\rho}\frac{\partial\rho}{\partial t}=-\nabla\cdot\mathbf{U}_{pol}
\label{continuity}%
\end{equation}
which can be combined with the induction equation to give%
\begin{equation}
\frac{\partial B_{\phi}}{\partial t}\simeq\frac{B_{\phi}}{\rho}\frac
{\partial\rho}{\partial t} \label{Bphiaccum}%
\end{equation}
showing that $B_{\phi}$ increases in \ proportion to the increase in mass
density at the stagnation layer. Since $I$ is constant and $2\pi rB_{\phi}%
=\mu_{0}I$ \ the current channel radius in the vicinity of the stagnation
layer must decrease as $B_{\phi}$ increases to keep $rB_{\phi}\ $constant.
Thus, the bulge of the current channel must diminish as sketched in Fig. 5
and, because $I=I(\psi)$, the bulge of the constant $\psi$ surfaces must also
diminish. The result is that the flux tube tends to become axially uniform,
hot, and dense.

\subsection{Changes in length}

Making the flux tube axially uniform increases $\mathbf{B}_{pol}$ because
squeezing the poloidal flux surfaces together results in a larger field. Since
$B_{pol}\sim B_{z}\ $ is much larger than $B_{\phi}$, it would seem that too
much energy would have to be invested into squeezing the poloidal flux
surfaces together. However, if we recall that the loop is really arched and
allow the loop length to change in such a way that $\int\mathbf{B}_{pol}\cdot
d\mathbf{l}$ remains constant where the line integral is over the length of
the loop, then the loop length $2h$ will become shorter as $\mathbf{B}_{pol}$
increases. If $\int\mathbf{B}_{pol}\cdot d\mathbf{l=}const\mathbf{.,}$ the
stored energy in the poloidal field is
\begin{align}
W_{pol}  &  =\frac{1}{2\mu_{0}}\int B_{pol}^{2}\ d\mathbf{l\cdot}%
d\mathbf{s}\nonumber\\
&  =\frac{1}{2\mu_{0}}\int\mathbf{B}_{pol}\cdot\ d\mathbf{l\ }\int
\mathbf{B}_{pol}\cdot d\mathbf{s}\nonumber\\
&  =\frac{\psi}{2\mu_{0}}\int\mathbf{B}_{pol}\cdot\ d\mathbf{l\ }\ \nonumber\\
&  =const. \label{W}%
\end{align}
It is reasonable to assume that $\int\mathbf{B}_{pol}\cdot d\mathbf{l}$
remains constant, because $\mathbf{B}_{pol}$ is produced by currents external
to the flux tube (e.g., by the subsurface currents sketched in Fig. 2a). These
source currents may be assumed to stay constant on the time scale during which
the flux tube undergoes stages 1-3. If one follows the poloidal field along
its entire length both above and below the solar surface, then it must satisfy
Ampere's law
\begin{align}
\mu_{0}I_{ext}  &  =\oint\mathbf{B}_{pol}\cdot d\mathbf{l}\nonumber\\
&  =\int_{loop}\mathbf{B}_{pol}\cdot d\mathbf{l+}\int_{subsfc}\mathbf{B}%
_{pol}\cdot d\mathbf{l} \label{subsfc}%
\end{align}
where the contour consists of the loop above the surface and, in addition, the
subsurface portion; the contour links links the subsurface source current
system denoted as $I_{ext}$. It seems reasonable to assume that the subsurface
field remains invariant during stages 1-3 and so $\int_{loop}\mathbf{B}%
_{pol}\cdot d\mathbf{l}$ must also remain constant.

Thus, as the poloidal field lines squeeze together to make the flux tube
axially uniform, the flux tube becomes shorter in such a way as to keep the
energy stored in the poloidal field constant. In this manner, no work needs to
be done to squeeze the poloidal flux surfaces together. One can imagine that
the \textquotedblleft field line tension" of the poloidal field shortens the
length of the loop as the poloidal field is made stronger when the poloidal
flux surfaces are squeezed together.

\subsection{Ultimate beta}

\qquad The $z-$directed force given by Eq.(\ref{Fz}) can be written as%
\begin{equation}
F_{z}=-\frac{\partial}{\partial z}\left(  \frac{B_{\phi}^{2}}{\mu_{0}}\right)
. \label{Fz2}%
\end{equation}
The quantity $B_{\phi}^{2}/\mu_{0}$ can be considered as an effective
potential energy and so the fact that $B_{\phi}^{2}/\mu_{0}$ is large at
$z=\pm h$ and small at $z=0$ means that there is an effective potential well
which the plasma falls down as it moves from $z=\pm h$ to $z=0$. The order of
magnitude of the resulting flow velocity is given by the reduction in
potential energy due to the plasma falling down the slopes of the well and so
the resulting flow velocity will be $U_{z}^{2}\simeq B_{\phi}^{2}/\mu_{0}\rho$
where $B_{\phi}^{2}$ is evaluated at $z=\pm h$ where $B_{\phi}^{2}$ is
largest. Thus, the flow velocity is of the order of the Alfv\'{e}n velocity
calculated using the toroidal field (this is much smaller than the Alfv\'{e}n
velocity calculated using the poloidal field on the assumption that the flux
tube is only slightly twisted). At the stagnation layer the converging flow
velocity is thermalized and so the plasma pressure at the stagnation layer
will be of the order of $P=nm_{i}v_{Ti}^{2}+nm_{e}v_{Te}^{2}\simeq\rho
U_{z}^{2}=\left[  B_{\phi}^{2}/\mu_{0}\right]  _{z=\pm h}.$ Assuming that
$B_{\phi}<<B_{pol}$, the plasma $\beta$ resulting from flow stagnation is
therefore
\begin{equation}
\beta\simeq\frac{2\mu_{0}P}{B_{pol}^{2}}=2\frac{B_{\phi}^{2}}{B_{pol}^{2}%
}=\left(  \frac{\mu_{0}I}{\psi\ }\right)  ^{2}\frac{a^{2}}{2\ } \label{beta2}%
\end{equation}
where $a$ is the radius of the flux tube and $\psi=B_{pol}\pi a^{2}.$ Using
the definition $\alpha=\mu_{0}I/\psi$ then
\begin{equation}
\beta=\alpha^{2}a^{2}/2 \label{betacondition}%
\end{equation}
is the value of $\beta$ resulting from flow stagnation.

The diminishing of the bulge squeezes together the poloidal field so that
there will be a finite $J_{\phi}$, but if the flux tube is squeezed to the
point of being axially uniform, then $J_{\phi}$ vanishes again. Thus $J_{\phi
}$ starts out by being zero, becomes finite, and then becomes zero again if
and when the flux tube becomes straight.

If an equilibrium is established then $\mathbf{J\times B}=\nabla P\ $which
implies $\mathbf{B\cdot}\nabla P\mathbf{=}0$ and $P=P(\psi).$ We define
$\psi_{0}$ as the flux surface on which $P$ vanishes and $\bar{\psi}%
(r,z)=\psi(r,z)/\psi_{0}$ as the normalized flux so that
\begin{equation}
P(\psi)=(1-\bar{\psi})P_{0} \label{P-profile}%
\end{equation}
where $P_{0}$ is the on-axis pressure. The equilibrium equation
$\mathbf{J\times B}=\nabla P$ can\ then be written in
Grad-Shafranov\cite{Grad-Shafranov} form as
\begin{equation}
r\frac{\partial}{\partial r}\left(  \frac{1}{r}\frac{\partial\bar{\psi}%
}{\partial r}\right)  +\frac{\partial^{2}\bar{\psi}}{\partial z^{2}}%
+\alpha^{2}\bar{\psi}=2\frac{r^{2}}{a_{0}^{2}}\frac{\beta}{a_{0}^{2}%
}\ \label{GS}%
\end{equation}
where $a_{0}$ is the flux tube radius at $z=0$ and $\beta$ is defined in terms
of the mean axial field at $z=0,$ i.e., $\beta=2\mu_{0}P_{0}/\left(  \psi
_{0}/\pi a_{0}^{2}\right)  ^{2}.$ The general solution to Eq.(\ref{GS}) is
\begin{equation}
\bar{\psi}(r,z)=2\frac{r^{2}}{a_{0}^{2}}\frac{\beta}{\ \ \alpha^{2}a_{0}^{2}%
}+\gamma\chi(r,z) \label{GSsoln}%
\end{equation}
where $\chi(r,z)$ is any solution to the homogeneous equation%
\begin{equation}
r\frac{\partial}{\partial r}\left(  \frac{1}{r}\frac{\partial\chi}{\partial
r}\right)  +\frac{\partial^{2}\chi}{\partial z^{2}}+\alpha^{2}\chi
=0\ \label{chieq}%
\end{equation}
and $\gamma$ is a constant to be determined. The condition $\bar{\psi}(r,z)=1$
when $z=0$ and $r=a_{0}$ fixes $\gamma$ so that the general solution to the
Grad-Shafranov equation is thus
\begin{equation}
\bar{\psi}(r,z)=\ \frac{2\beta r^{2}}{\ \alpha^{2}a_{0}^{4}}+\left(
1-\frac{2\beta}{\ \ \alpha^{2}a_{0}^{2}}\right)  \frac{\chi(r,z)}{\chi
(a_{0},0)}. \label{GSsoln2}%
\end{equation}
If $\beta=\alpha^{2}a_{0}^{2}/2$ then the \textit{only} solution to
Eq.(\ref{GS}) satisfying the prescribed boundary condition that $P$ vanishes
when $\psi=\psi_{0}$ is the particular solution $\psi=\psi_{0}r^{2}/a_{0}%
^{2}.$ However, this solution is axially uniform and so we no longer need to
specify $a_{0}$ as being the radius at $z=0;$ it is in fact the radius at all
$z.$ Equation (\ref{GSsoln2}) provides the important result that having a
finite but extremely small $\beta$ will cause the poloidal flux surfaces to
differ substantially from the force-free situation where $\beta$ is exactly
zero and in particular will cause the system to become axially uniform when
$\beta=\alpha^{2}a_{0}^{2}/2.$ From a mathematical point of view this is
because the right hand side of Eq.(\ref{GS}) is an inhomogeneous term (source
term) of the partial differential equation. The source term results in there
being a particular solution which would not exist if Eq.(\ref{GS}) were
homogeneous, i.e., if $\beta$ were exactly zero.

The axial uniformity condition $\beta=\alpha^{2}a^{2}/2$ is just \ Eq.
(\ref{betacondition}) and so we conclude that the $\beta$ produced by flow
stagnation is precisely the $\beta$ required to force the Grad-Shafranov
equation to give an axially uniform solution. We further conclude that, given
sufficient time and assuming there are no losses, current-carrying flux tubes
will always tend to become axially uniform and will always tend to have the
$\beta$ given by Eq.(\ref{betacondition}). This result is similar to the
well-known property \cite{Jensen:2002} of tokamaks having $\beta_{pol}$ of
order unity because diamagnetism exactly balances paramagnetism so that the
resulting field is a potential (vacuum) field. The roles of poloidal and
toroidal directions are interchanged in the coronal loop compared to a tokamak
and so in the coronal loop it is $\beta_{\phi}$ which is of order unity.

The predicted $\beta$ can be compared with actual observed values of $\beta$
in solar coronal loops. To make a prediction, a nominal observed flux loop
radius $a=1.6\times10^{6}$ m \cite{Aschwanden:2000} and a nominal measured
active region $\alpha=2\times10^{-8}$ m$^{-1}$ are used \cite{Pevtsov:1997}.
These parameters predict a nominal $\beta_{predicted}=\alpha^{2}%
a^{2}/2=\allowbreak5\times10^{-4}.$ The observed value $\beta_{observed}$ is
calculated using a nominal density $n=10^{15}\,\ $m$^{-3},$ and a nominal
temperature $10^{6}$ K \cite{Aschwanden:2000}. In addition a nominal axial
magnetic field $B_{z}=1.5\times10^{-2}\,$\ T is assumed based on the argument
that since the flux tube is axially uniform, its axial field must also be
axially uniform and so will have the same value as the nominal $B_{z}%
=1.5\times10^{-2}\,$\ at the surface of an active region. These parameters
give $\beta_{observed}=2\mu_{0}n\kappa T/B_{z}^{2}=4\times10^{-4}$ which is
similar to the predicted value.

This model also has implications regarding the brightening typically observed
when the axis of a coronal loop starts to writhe and the loop develops a kink
instability (sigmoid). Since kink instability occurs when $\alpha h\sim2\pi$
\cite{Kruskal,Hood:1979,Linton:1996,Rust:1996,Hsu:2002} and for a long thin
flux tube $a<<h$, this model predicts that $\beta=\alpha^{2}a^{2}/2<<$
$\alpha^{2}h^{2}/2$ will still be small even if $\alpha$ is increased to the
point where $\alpha h\sim2\pi$ and kink instability occurs. However, $\beta$
will increase as $\alpha$ increases. Since the brightness of a loop is
proportional to $n^{2}$ for a given temperature, this model predicts that the
loop should brighten in proportion to the writhing of its axis (i.e., in
proportion to $\alpha$ as $\alpha h$ approaches unity).

The model thus provides a heating mechanism (stagnation of MHD-driven flows)
which is consistent with observed coronal temperatures and densities; however,
the prediction is for $\ n\kappa T$ rather than for temperature or density
separately; a more detailed analysis would be required to isolate the
individual dependence of temperature and density on the stagnation process.

\subsection{Energetic Tail}

As the flows converge, there will be a few particles which have collision mean
free paths and trajectories such that they bounce back and forth between
converging fluid elements. Because these particles gain energy on each bounce
between the converging flows, these particles will gain energy without bound
until desynchronized or lost, i.e., they will undergo Fermi acceleration
\cite{Uchida:1988}. The number of particles having the appropriate mean free
path will be small, so one will expect a small high energy tail located around
$z=0.$ The concentration of high energy particles around $z=0,$ i.e., at the
top of a loop, is in fact what is observed \cite{Feldman:2002}.

\section{Summary and conclusions}

We have shown that the apparently simple problem of driving an electric
current along a pre-existing potential magnetic flux tube is actually quite
complicated and consists of three stages. In real situations these stages
would overlap and not be as distinct as outlined here.

The first stage involves a twisting of the magnetic field and an associated
$z$-dependent toroidal rotation (i.e., twisting)\ of the plasma; this motion
is incompressible. The second stage involves convergent axial flows driven by
the nonlinear, non-conservative force associated with the axial gradient of
$B_{\phi}^{2}/\mu_{0}$. The third stage involves accumulation of both mass and
toroidal flux at $z=0$ and a simultaneous conversion of directed flow energy
into thermal energy, i.e., stagnation. The concomitant increase in toroidal
magnetic field at the stagnation layer ultimately leads to an equilibrium and
because the flow stagnation gives $\beta=\alpha^{2}a^{2}/2$ the equilibrium is
axially uniform. \ This sequence of events should be quite common and should
explain why current-carrying flux tubes are so often observed to be filamentary.

Finally, it should be emphasized that symmetries in both $\phi$ and in $z$
play a critical role in the behavior described here. Symmetry in $\phi$ (i.e.,
toroidal symmetry) prevents the existence of any toroidal electrostatic field
so that the only allowed toroidal electric field is the toroidal electric
field associated with changing poloidal flux, i.e., $E_{\phi}=-(2\pi
r)^{-1}\partial\psi/\partial t.$ Particles are therefore constrained to stay
within a poloidal Larmor radius of a flux surface so that there can only be AC
currents in the direction normal to a flux surface in which case the plasma
acts like a capacitor in the direction normal to the flux surfaces. Because
toroidal acceleration is driven only by the current normal to a flux surface
and because no toroidal pressure gradient is allowed, the toroidal motion is
constrained to be transient, finite, and dependent on the temporal behavior of
$I.$ Thus when $I$ is constant, poloidal current flows along poloidal flux
surfaces and there is no toroidal motion. Symmetry about the $z=0$ plane
causes this plane to be a stagnation layer where opposing plasma jets collide
resulting in accumulation of mass and of frozen-in toroidal magnetic flux and
also a thermalization of the flow kinetic energy.

\bigskip

Acknowledgement:\ Supported by United States Department of Energy Grant DE-FG03-97ER544.

\pagebreak

\bigskip

\end{document}